\documentclass[12pt,preprint]{aastex}

\usepackage{lscape}
\newcommand\hour{\mbox{$^{\mathrm h}$}}
\newcommand\mint{\mbox{$^{\mathrm m}$}}

\shorttitle{3D spectroscopy in BCGs}
\shortauthors{Mart\'{\i}nez-Delgado et al.}

\begin{document}

\title{3D Spectroscopy of Blue Compact Galaxies. Diagnostic Diagrams}
\author{Ismael Mart\'{\i}nez-Delgado\altaffilmark{1} ,  
Guillermo Tenorio-Tagle\altaffilmark{2}, Casiana Mu\~noz-Tu\~n\'on\altaffilmark{1},
Alexei V. Moiseev\altaffilmark{3}}
\and
\author{Luz M. Cair\'os\altaffilmark{4}}
\affil{Instituto de Astrof\'{\i}sica de Canarias, V\'{\i}a
 L\'actea, E-38200 La Laguna, Tenerife, Canary Islands, Spain;
       \email{idelgado@iac.es, cmt@iac.es}}
\affil{Instituto Nacional de Astrof\'{\i}sica, \'Optica y Electr\'onica, AP 51, 72000 Puebla, 
       Mexico;\email{gtt@inaoep.mx}}
\affil{Special Astrophysical Observatory, 369167 Nizhnij Arkhyz, Russia;
\email{moisav@sao.ru}}
\affil{Astrophysikalisches Institut Postdam, An der Sternwarte 16, D-14482 Potsdam, Germany;
\email{luzma@aip.de}}

\begin{abstract}

Here we present the analysis of 3D spectroscopic data of three Blue Compact Ga\-la\-xies (Mrk~324, Mrk~370, and 
III~Zw~102).
Each of the more than 22500 spectra obtained for each galaxy has been fitted  by a single gaussian  from which we have inferred 
the velocity dispersion ($\sigma$), the peak intensity  (I$_{peak}$), and the 
central wavelength ($\lambda_c$). The analysis shows that the 
$\sigma$ {\it vs} I$_{peak}$ diagrams look remarkably similar to those 
obtained for giant extragalactic H {\small II} regions. 
They all present a supersonic narrow horizontal band that extends across  all the 
range of intensities and that result from the massive nuclear 
star-forming regions of every galaxy. The $\sigma$ {\it vs} I$_{peak}$ diagrams  present also
several inclined bands of lower intensity and an even larger $\sigma$, 
arising from the large galactic volumes that  surround the 
main central emitting knots.
Here we also show that the $\sigma$ 
{\it vs} $\lambda_c$ and $\lambda_c$ {\it vs} I$_{peak}$ diagrams, are powerful tools able to unveil 
the presence of high and low mass stellar clusters, and thus allow for 
the possibility of inferring the star formation activity of 
distant galaxies, even if these are not spatially resolved.

\end{abstract}


\section{Introduction}

High resolution panoramic spectroscopy with good spatial and spectral resolution is known to be a powerful tool for 
studying the kinematics 
of ionized nebulae as it leads to a simultaneous mapping, at seeing limited resolution, of a 
particular emission line over the whole nebula.
This however can easily  lead to several tens of thousands of spectra, making it difficult 
to issue a detailed 
and/or a global interpretation  of the data.  
An analysis procedure that has proven to be simple and powerful results from fitting a single 
gaussian to each of the resultant 
emission line profiles \citep[see][]{Cas94,Cas95}. A single gaussian regardless of the actual degree of asymmetry 
or splitting in the line profiles. The fit is to
conserve the flux of the line profile, and thus lower intensity but broader lines would 
result from the most asymmetric or largely
splitted line profiles.   
From the resultant fits one can derive the velocity dispersion ($\sigma$), the peak 
intensity (I$_{peak}$) of the fitted lines
as well as their central wavelength ($\lambda_c$). 
Such a method when applied to  giant extragalactic H {\small II} region (GHIIR) data, leads to two 
distinct regions in the 
$\sigma$ {\it vs} I$_{peak}$ diagram
\citep[see][]{Cas96}: A  supersonic ($\sigma$ $>$ c$_{\mathrm H\, \mbox{\tiny II}}$), 
relatively narrow horizontal band with all possible peak intensity values, and a second region populated by   
lower intensity points presenting even larger supersonic $\sigma$ values, that crowd along 
multiple inclined 
bands. For the case of GHIIRs it has been shown 
\citep[see][the last reference for the case of H {\small II} Galaxies data]{Cas94, Cas96, Nancy95, Oriol00, Telles01}  
that the horizontal band 
is conformed by data with truly gaussian profiles arising mainly from
the brightest portions of the nebula, from the small regions that enclose groups or clusters 
of stars. The inclined lower intensity 
and highly supersonic bands on the other hand,  emanate from multiple photoionized shells that surround the star-forming 
centers and are likely to 
result from the stellar  mechanical energy impact into the 
ISM, and thus from single gaussian fits to highly asymmetric or strongly splitted
lines.
The analysis of the $\sigma$ {\it vs} I$_{peak}$ diagram for GHIIRs in nearby 
galaxies (closer than 1 Mpc) was then proposed as an excellent tool to determine their degree of evolution on 
kinematic bases \citep{Cas96}. 
Here we extend the analysis to Blue Compact Galaxies (BCGs) at even larger distances and show that they experience a 
similar dynamical evolution (ie. their $\sigma$ {\it vs} I$_{peak}$ diagrams look very much like those of GHIIRs). 
We further extend our analysis by looking also at the $\sigma$ {\it vs} $\lambda_c$ and $\lambda_c$ {\it vs} 
I$_{peak}$ diagrams, and show them as ideal tools to distinctly unveil the presence of high and low mass stellar clusters.

Here  section 2 provides a short description of the observations. Section 3 contains a detailed analysis of the data 
by means of the diagnostic diagrams here proposed and section 4 gives a summary and a discussion of our findings.

\section{The observations}
We use the 6m Big Telescope Alta-azimuthal (BTA) of the Special Astrophysical Observatory 
(SAO) in Nizhnij Arkhyz (Russia), equipped with
the multi-mode focal reducer SCORPIO\footnote[1]{http://www.sao.ru/hq/lsfvo/devices/scorpio/scorpio.html} 
in the scanning Interferometer Fabry-Perot (IFP) observational mode  
\citep[see][]{Afanasiev05}, for the observations of three BCGs:  Mrk~370, III~Zw~102, and Mrk~324.  SCORPIO is 
provided with a EEV 42-40 CCD array of 2048 $\times$ 2048 pixels with 
an instrumental scale of 0.18 $''$ pixel$^{-1}$. A set of interference filters centered at the H$\alpha$ line,
covering an interval of systemic velocities that range from $-$200 to 10000 km s$^{-1}$,  is 
used at the IFP observational mode. The IFP offers two configuration depending on the interference order, in our case
at the H$\alpha$ line, the IFP260 (at the interference order 235) and the IFP500 (at the interference order 501). 
Our observation runs were performed using the IFP500 mode, that offers a better spectral resolution in the Ha 
emission line.

The output from the IFP results on a 3D data cube with {\it x} and {\it y} being the
spatial plane and {\it z}  the wavelength sampling or ethalon step, that in the case of IFP500 mode, it 
translates into 32-40 channels. The original data cube dimensions were
(522,522,36) although for our analysis we reduced the size, matching the objects size, to (150,150,36).
The Free Spectral Range (FSR) was 13 \AA, enough to sample the H$\alpha$ line, that
was scanned with a spectral sampling of 0.7 \AA\ $\approx$ 32 km s$^{-1}$. A  
binning of 4$\times$4 pixels was applied in the spatial directions in order to reduce the time exposure and 
improve the signal to noise of the final data cube, resulting in a final spatial scale of 0.7 $''$
pixel$^{-1}$. The instrumental width taken from a fitting of the calibration lamp data cube was
$\sigma_{inst}=20\pm3$ km s$^{-1}$. Table \ref{logs} lists a log of the observations together with 
some data about the galaxies:  Column 1 gives the name of the object; columns 2 and 3 show the coordinates 
(from NED\footnote[2]{NASA/IPAC Extragalactic 
Database is operated by the Jet Propulsion Laboratory, California Institute of Technology, under 
contract with the National Aeronautics and Space Administration(NASA).}); columns 4 shows the linear scale for
each galaxy; column 5 lists the
total time exposure of the observations; column 6 gives the seeing conditions; column 7 shows the
systemic velocity of each galaxy and column 8 gives the distance in Mpc.
Each of our SCORPIO trimmed data cubes contains a total of 22500 spectra from an area of $\sim$3 arcmin$^2$. 
The bulk of the data thus requires of automatic techniques of analysis to extract the full power of 
3D spectroscopy. For each of the three galaxies, their H${\alpha}$ emission lines in every 
pixel were fitted by single gaussians, to then obtain their I$_{peak}$, $\sigma$, and 
$\lambda$c, as described in \cite{Cas95, Cas96}. The velocity dispersion is given in km s$^{-1}$ and 
is corrected for the instrumental and thermal broadening 
($\sigma^2=\sigma^2_{obs}-\sigma^2_{inst}-\sigma^2_{th}$). 
We then used {\it MoisSoft}, the software package  designed 
for the manipulation of SCORPIO-IFP spectral line data cubes. For a detailed explanation about the data 
reduction procedure, see \cite{Alexei02}.

\section{Three Dimensional  Spectroscopy of BCDs}

\subsection{Mrk~324}

Mrk~324 is a galaxy classified as a Blue Compact Dwarf \citep[BCD, see][]{Thuan81} and is included as a Nucleated BCD
in the morphological classification of   Cairos et al. (2001), as it is  basically  powered by a luminous 
central knot. The star formation history of the galaxy 
fits well with an instantaneous star formation law with a Salpeter IMF ($\alpha=2.35$) and a range of masses between 
1$\leq{M/M_{\odot}}\leq$100, in which a central knot of 4.4 Myr and 1.4$\times10^5$ M$_\odot$ in stars 
dominates  95\% of the emission flux of the galaxy \citep[see][]{Ismael06}.

\begin{landscape}
\begin{figure}
\begin{center}
\includegraphics[width=22cm]{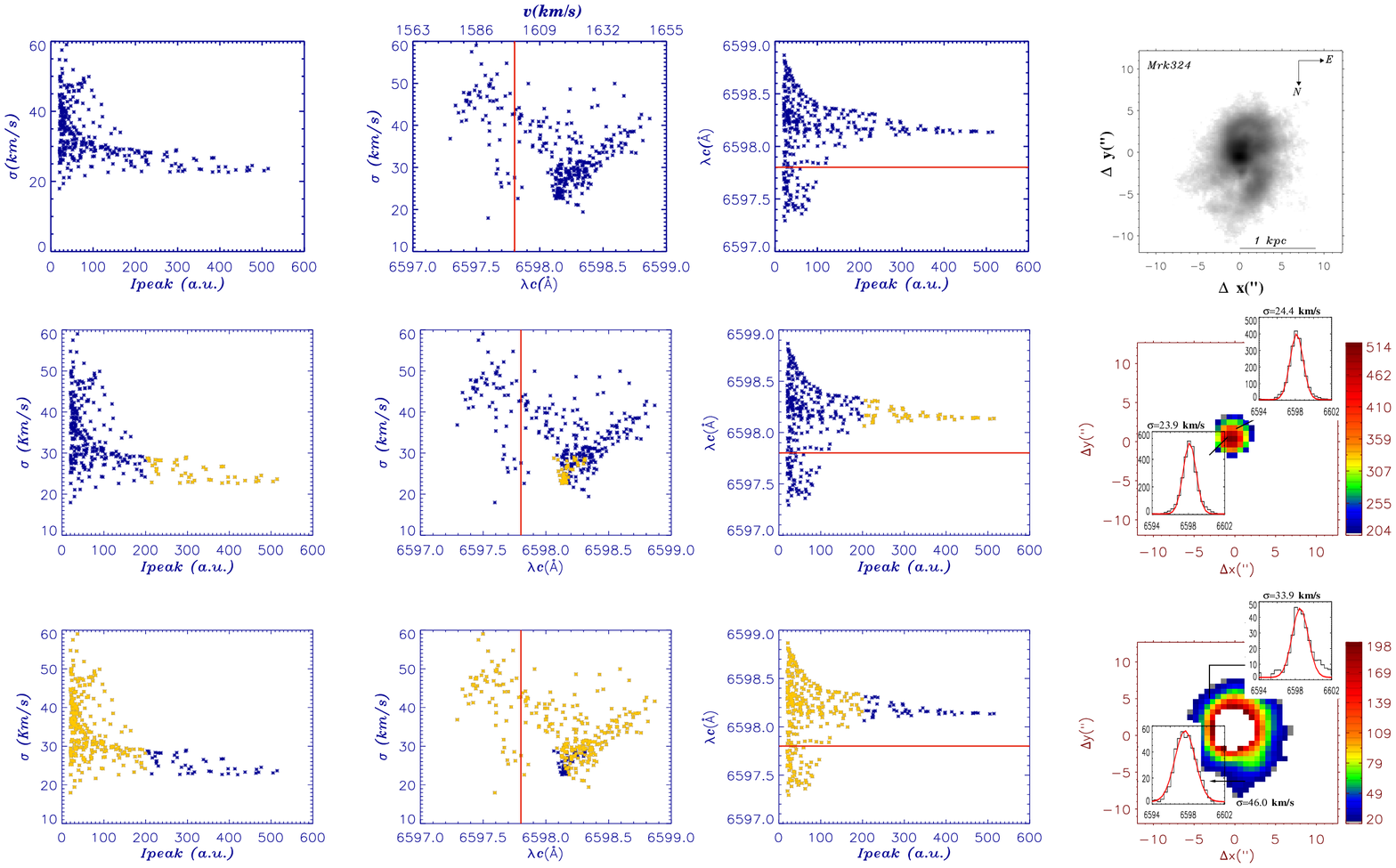}
\caption{\footnotesize The figure shows in the first row of panels the velocity dispersion $\sigma$ {\it vs} 
peak intensity (the $\sigma$ was corrected for the instrumental and thermal broadening), for all the 
gaussian fits to the individual spectra of Mrk~324. The second and third panels display the 
corresponding $\sigma$ {\it vs} central lambda 
($\lambda_c$, indicating also the velocity range in km s$^{-1}$) and $\lambda_c$ {\it vs} peak 
intensity, respectively. The solid line in these two panels indicates the rest velocity of the galaxy.
The last panel shows the H$\alpha$ image of the galaxy and indicates its scale in kpc. 
The second and following rows of panels highlight in colour 
the various sets of data points selected in the $\sigma$ {\it vs} peak intensity 
diagram and their location in the $\sigma$ {\it vs} $\lambda_c$ and  $\lambda_c$ {\it vs} peak 
intensity planes. The last panels in every row, with a 
peak intensity scale,  
display the locations in the galaxy that produce the selected data points.  Inserts in these panels
correspond to typical gaussian fits (solid red lines) to the data arising
from different regions (axis correspond to intensity, in counts,  vs the wavelength in \AA). The 
fitted $\sigma$ values are indicated in every frame. \label{Mrk324diag}}
\end{center}
\end{figure}
\end{landscape}

Figure \ref{Mrk324diag} (upper row of panels) shows the results from the single gaussian fit to the full 
H$\alpha$ SCORPIO data set in the $\sigma$ {\it vs} I$_{peak}$,  $\sigma$ {\it vs} $\lambda_c$, and $\lambda_c$ 
{\it vs} I$_{peak}$ diagrams for Mrk~324.
The last frame, upper row, shows the H$\alpha$ image of the nucleated galaxy \citep[see][]{Cai01b} 
obtained at the NOT 2.52m telescope of the Roque de Los Muchachos Observatory (ORM) in La Palma (Spain).
The resemblance  of the  $\sigma$ {\it vs} I$_{peak}$ diagram with those of GHIIRs is remarkable. 
It presents a  horizontal band that limits the values of $\sigma$ to supersonic values 
($\sigma$ $\geq$ 20 km s$^{-1}$)
and a bunch of low intensity points at even higher $\sigma$ values, that delineate a 
triangular structure above the 
horizontal band. If one selects from the $\sigma$ {\it vs} I$_{peak}$ diagram 
the highest intensity data points, those belonging to the  
horizontal band (see second row of panels in Figure \ref{Mrk324diag}), these in the $\sigma$ {\it vs} $\lambda_c$ 
diagram appear clumped at a given $\lambda_c$ ($\sim$ 6598.2 A) what implies that 
they all arise from a very similar location in the galaxy, that they all belong in principle 
to a single entity. Note however, that this is not exactly coincident with the rest wavelength of the galaxy
(marked in the $\lambda_c$ \ plots with a red line, see Figure \ref{Mrk324diag}). The selected points in the 
$\lambda_c$ {\it vs} I$_{peak}$ diagram defined also a pointed structure towards high 
intensity values. These points in the emission-line peak intensity map 
(last frame) display their location in the galaxy, which indeed coincides 
with the major burst of star formation in Mrk~324.   A major burst with an estimated
mass in stars $\sim 4 \times 10^5$ M$_\odot$ and an age of about 4.4 Myr \citep[see][]{Ismael05,Ismael06}
that causes an H {\small II} region with a radius of $\sim$ 250 pc.
On the other hand, the data belonging to the inclined bands with a higher velocity 
dispersion and a lower intensity in the $\sigma$ {\it vs} I$_{peak}$ diagram  
(see third row of panels in Figure \ref{Mrk324diag}) appear well scattered in $\lambda_c$ and arise 
from a large-scale (600 pc in thickness) expanding rim that surrounds the
central emitting knot, as shown in the last panel of the third row.
The inserts in the last column of Figure \ref{Mrk324diag} display the H$\alpha$ line and its single gaussian
fit at various locations in the galaxy.
As in GHIIRs, the data corresponding to the horizontal band in the $\sigma$ {\it vs} I$_{peak}$ 
diagram leads to the highest intensity true gaussian profiles, while the profiles fitted to lower intensity 
and larger $\sigma$ data points, arising from the outer expanding rim, are fits to asymmetric and/or double
peak line profiles. Figure \ref{Mrk324diag} shows the quality of typical line profiles from each of the
regions, displaying their intensity, central wavelength, and $\sigma$ values. 

\subsection{Mrk~370}

This BCD galaxy belongs to the Chained morphological  type \citep[see][]{Cai01b}, in which the star formation 
takes place along a privileged direction. It follows the same star formation law as Mrk~324, with 25 young and massive exciting clusters 
hosted over an underlying stellar component with ages larger than 5 Gyr, (\citealp[see][]{Cai02}). The clusters have 
an average age of 6.6 Myr with a dispersion of 0.9 Myr, suggesting a coeval and global star formation event.  For the ones located in the inner nuclear region,  
the stellar mass content reaches up to several 10$^5$ M$_\odot$, 
\citep[see][]{Ismael06}.

\begin{landscape}
\begin{figure}
\begin{center}
\includegraphics[width=19cm]{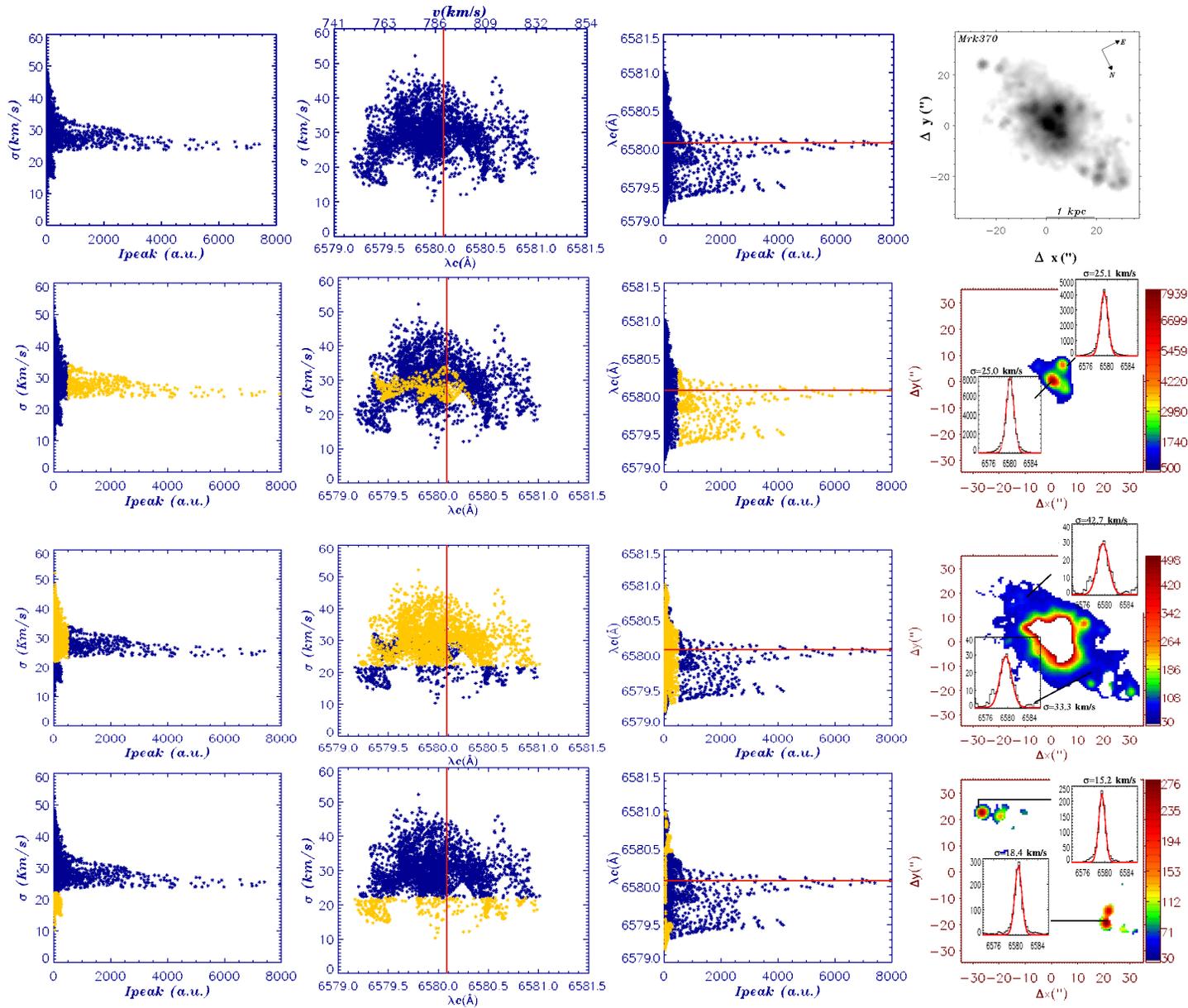}
\caption{Mrk~370. The same as Figure \ref{Mrk324diag} for Mrk~370. \label{Mrk370diag}}
\end{center}
\end{figure}
\end{landscape}

Figure \ref{Mrk370diag} displays similar diagnostic diagrams  for Mrk~370.
The last frame in the upper row of panels shows the H$\alpha$ image of the chained galaxy \citep[see][]{Cai01b} 
obtained at the CAHA 2.2m telescope of the Centro Astron\'omico Hispano Alem\'an (CAHA), from Calar Alto 
Observatory (Spain). 
The ionized gas in Mrk~370 (see the H$\alpha$ image in Figure \ref{Mrk370diag})
presents much more structure than Mrk~324. However, its    $\sigma$ {\it vs} I$_{peak}$ 
diagram looks almost identical to that shown in Figure \ref{Mrk324diag} for Mrk~324. The only
exception is perhaps the low intensity data points with values of $\sigma$ 
below the supersonic lower limit imposed by the horizontal band ($\leq $ 21 km s$^{-1}$, see first panel 
in Figure \ref{Mrk370diag}). 
The structure of the photoionized gas in Mrk~370 begins to be 
disentangled when one selects the data from the horizontal band in 
the $\sigma$ {\it vs} I$_{peak}$ diagram (see Figure \ref{Mrk370diag}, second row of panels). Such a data set 
appears clumped at 
three different $\lambda_c$ values in the $\sigma$ vs $\lambda_c$ diagram. The brightest one sits at the rest
wavelength of the galaxy (6580.1\AA). There is a second one at $\sim$6579.4\AA\ and a third one between these 
two at $\sim$6579.8\AA. 
Two of them are much more apparent in the $\lambda_c$ 
{\it vs} I$_{peak}$ diagram where they clearly end up as pointed structures 
with different intensity (factors of 2) and different 
central wavelengths. The emission-line peak intensity map tracing  the selected points (last frame in the second
row) shows two major bursts of stellar formation and two adjacent much less intense, 
associated 
to the massive central stellar clusters with masses that range between (8$-$50)$\times 10^4$ M$_\odot$ 
\citep[see][]{Ismael05,Ismael06}, also resolved by \cite{Cai02} using the INTEGRAL spectrograph 
with a single gaussian fit to the emission lines. 
All of these, although near the center of the galaxy, are
 kinematically resolved as well  detached structures. This multiplicity of structures would have 
been missed if one had only use the 
$\sigma$ {\it vs} I$_{peak}$ diagram  as a diagnostic for this galaxy (see first panel in  Figure \ref{Mrk370diag}),
as the brightest points present almost the same supersonic gaussian profiles.  These massive regions 
also match the area of the galaxy that shows the highest ionization values 
(see the H$\alpha$ emission-line and the [O{\small III}]/H$\beta$ maps in \citealp{Cai02}).

The third row of panels in Figure \ref{Mrk370diag} displays the data from  the inclined 
bands with low intensity and large $\sigma$ values in the
$\sigma$ {\it vs} I$_{peak}$ diagram.   This data set, as the corresponding one 
in Mrk~324, presents a large range of values of $\lambda_c$, around the systemic velocity of the galaxy.
The spread of the data is also noticeable in the $\lambda_c$ {\it vs} I$_{peak}$ plane. The bulk of the data
arises from  a large galactic volume shown in the emission-line peak intensity map as an expanding 
rim of gas around the main 
centers of star formation all over the galaxy. This is most intense around the 
nuclear starburst zone and although less intense, it is still noticeable around smaller bursts of 
stellar formation some 2 kpc away from the nuclear region. 

The final row of panels show the low intensity and low velocity dispersion data 
from the $\sigma$ {\it vs} I$_{peak}$ diagram (see Figure \ref{Mrk370diag}), which 
appears as  vertical well detached low $\sigma$ structures in the $\sigma$ {\it vs} $\lambda_c$ plane. 
These structures, given their low values of $\sigma$, detach well from the 
bulk of the data, most of which presents supersonic values, and due to their largely different 
$\lambda_c$ values they also appear as well detached entities along the $\lambda$ scale. 
Note however that  given their low intensity values,  they are much 
less apparent 
in the $\lambda_c$ {\it vs} I$_{peak}$ diagram.  These data points
correspond in the emission-line peak intensity map (last panel in the 4$^{th}$ row of Figure \ref{Mrk370diag})
to multiple knots of emitting gas, or  H {\small II} regions excited by low mass stellar groups 
\citep[with masses $\le$ 10$^4$ M$_\odot$, see][]{Ismael05,Ismael06} at
large distances from the centre of the galaxy. The line profiles arising from these regions 
are also gaussian profiles although with a very low intensity when compared to 
those arising from the central starburst regions (see inserts in the last frame of Figure \ref{Mrk370diag}).

\subsection{III~Zw~102}

This galaxy is believed to be the result of
an interaction event \citep[see][]{Vor59,Vor77} and has been considered by  some authors  
to belongs to the polar-ring galaxy type \citep{Whitmore90}. It is morphologycally considered as 
an Extended Blue Compact Galaxy \citep{Cai01b}, in which 
65 star-forming regions were identified spreaded over the
whole main body of the galaxy, including the arms. The age of the knots and their low age dispersion 
(6.1 $\pm$ 0.6 Myr) suggest a coeval starburst event \citep[see][]{Ismael06}.

\begin{landscape}
\begin{figure}
\begin{center}
\includegraphics[width=20cm]{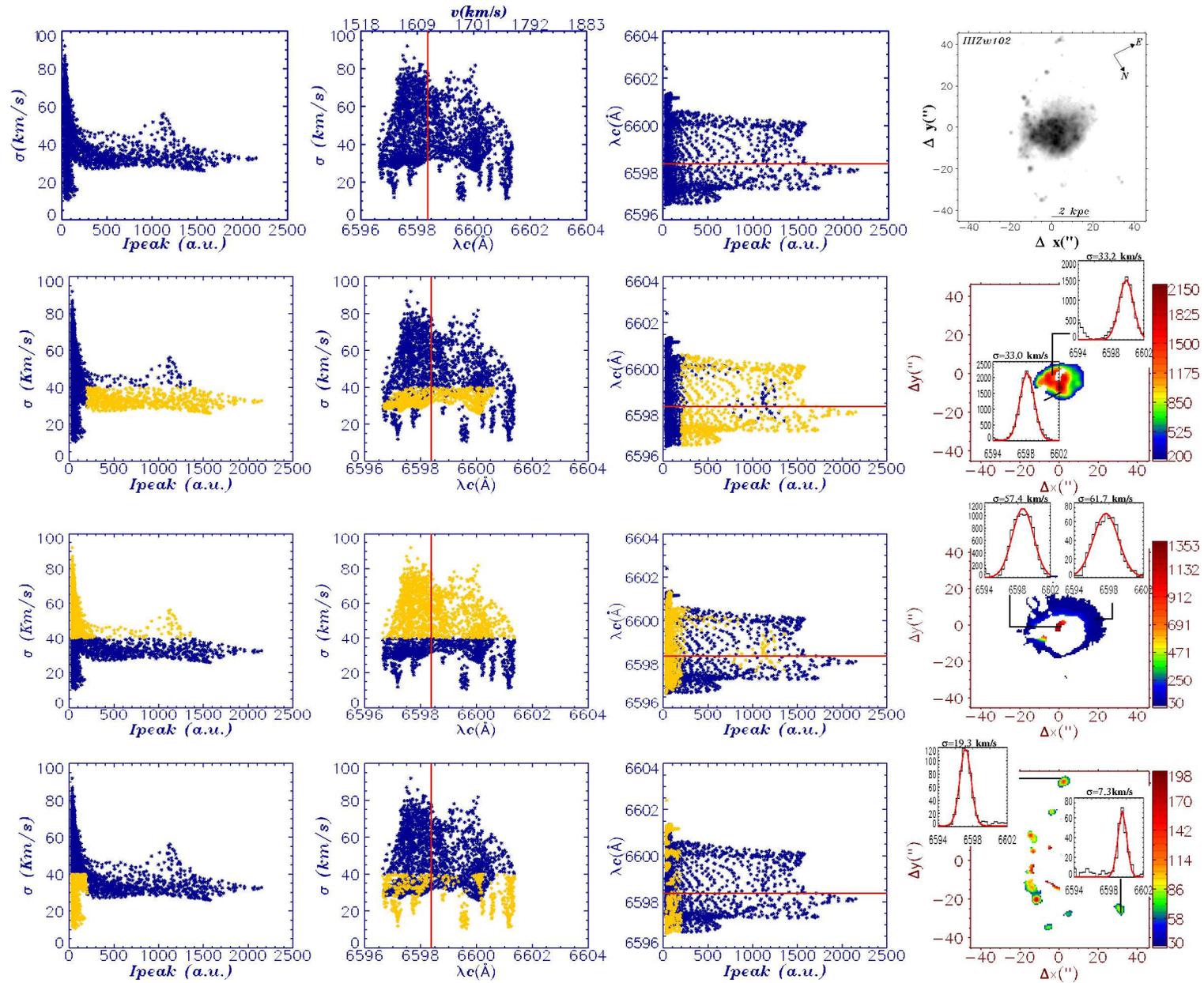}
\caption{III~Zw~102. The same as Figure \ref{Mrk324diag} for III~Zw~102.\label{IIIZw102diag}}
\end{center}
\end{figure}
\end{landscape}

Figure \ref{IIIZw102diag} displays the results for  III~Zw~102.
The last frame in the upper row of panels, shows the H$\alpha$ image of the extended galaxy \citep[see][]{Cai01b} 
obtained at the NOT 2.52m telescope of the Roque de Los Muchachos Observatory (ORM) in La Palma (Spain).

The data for III~Zw~102 confirms the power of all of the diagnostic diagrams here used. 
The galaxy has a plethora of small H {\small II} regions 
spread along two arms as well as several massive centers of stellar 
formation in the densest nuclear region 
with masses between $\sim$ $10^5-10^6$ M$_\odot$, \citep[see][]{Ismael05,Ismael06}. All of these,
as well as the large galactic volumes undergoing a supersonic expansion leave 
their signature in the diagnostic diagrams.
In this way, the horizontal band with supersonic $\sigma$ values at all possible 
intensities noticeable in the $\sigma$ {\it vs} I$_{peak}$ diagram
(see Figure 3, 2$^{nd}$ row of panels) is spread over a large range of $\lambda_c$ 
values in the $\sigma$ {\it vs} $\lambda_c$ diagram and into at least
four pointed structures (at $\lambda_c $ = 6596.7\AA, 6597.25\AA, 6598\AA, and 6600\AA) of 
different intensities in 
the $\lambda_c$ {\it vs} I$_{peak}$ diagram. All of these in the emission-line peak intesity 
map are well resolved as giant structures 
within the galactic nuclear region.
On the other hand, the larger $\sigma$ points at all  peak intensities  in the 
$\sigma$ {\it vs} I$_{peak}$ diagram 
arise from gas expanding around the nuclear zone (see
third row of panels in Figure 3). Among this data set is the inclined structure between I$_{peak}$=1100-1400
which is produced by gas between the four major centers of stellar formation in the nuclear region. The broad
line structure appears at the center of the galaxy with a $\lambda_c$  very similar to the systemic
velocity of III~Zw~102.

Finally, the lowest $\sigma$ and lowest 
intensity data points   in the $\sigma$ {\it vs} I$_{peak}$ diagram
(last row of panels in Figure 3) appear, as in the case of Mrk370, as  a set of well detached vertical 
stalactites in the $\sigma$ {\it vs} $\lambda_c$ diagram, hanging from the bulk of the data at very distinct
$\lambda_c$ values. The latter as shown in the last panel are produced by 
the numerous small H {\small II} regions located along the 
galactic arms. As shown by the inserts in the last column of panels, the quality of the gaussian fits 
is totally different in the three regions. The nuclear zone shows supersonic perfect gaussians while the
surrounding gas presents much broader and asymmetric lines. Single gaussians with small $\sigma$ values fit
well the data from small centers of star formation.

\section{Summary}

Tridimensional spectroscopy with good spatial and spectral resolution, 
sampling a particular emission line over the whole area covered by an 
emitting nebula, is known to be the most
suitable observational technique for analyzing the global kinematics of GHIIRs. 
In particular if a single gaussian fit is carried 
out over the emission lines in every pixel and from those, one infers the velocity 
dispersion, the central wavelength, and the
peak intensity of the lines. The resultant  
$\sigma$ 
{\it vs} I$_{peak}$ diagram \citep{Cas96}
has been shown to be an excellent diagnostic diagram to separate the main broadening mechanisms 
affecting the emission lines, e.g., those that lead to shells and loops generated by the   
violent action from massive stars (see Dyson 1979, Roy et al. 1986)
which finally lead to cloud dispersal, and the one(s) affecting the regions of massive star formation 
(either gravity; see Terlevich \& Melnick 1981, Tenorio-Tagle et al. 1993, or turbulence;
see Chu \& Kennicutt 1994). The sequences shown by the various GHIIRs so far analyzed, 
led the authors also to propose the $\sigma$-I$_{peak}$ plot as a tool able to trace the 
evolutionary status of a GHIIR, from its formation to the total dispersal of the 
ionized gas.

Here we have confirmed that the data from the inclined supersonic bands in the $\sigma$ {\it vs} I$_{peak}$
diagrams are caused by the mechanical energy deposited by massive stellar sources. These, depending on the
age of the stellar clusters and on the density of the surrounding gas may affect small or large galactic volumes. However in all cases,
the interaction leads to very asymmetric line profiles, evident in the $\sigma$ {\it vs} I$_{peak}$ diagram for
their highly supersonic $\sigma$ values.

Here we have extended the analysis to Blue Compact Galaxies. 
We have selected sources with a single nuclear starburst (as Mrk~324) as well as galaxies with 
multiple stellar bursts both within their 
nuclei as well as at large galactocentric distances. In all cases the $\sigma$ 
{\it vs} I$_{peak}$ diagram resembles those from GHIIRs in which the data define a supersonic 
$\sigma$ (larger than $c_{\mathrm H\, \mbox{\tiny II}}\approx$ 10 km s$^{-1}$) horizontal band with all possible 
intensity values. Note that the value of the limiting  $\sigma$ is different in every case. For  III~Zw~102 
the horizontal band peaks at $\sim33$ km s$^{-1}$, while for  Mrk~324 and Mrk~370 it occurs at
24 and 25 km s$^{-1}$, respectively.

The $\sigma$ horizontal band, for the case of multiple massive nuclear bursts 
of stellar formation, has been shown to split into different bands in the $\lambda_c$ 
{\it vs} I$_{peak}$ diagram. In this new diagram, the largest nuclear regions of massive star formation
become evident due to the good spectral resolution. And thus even
with a poor spatial resolution, or when dealing with further away sources, the sampling of the velocity field, 
with our 3D spectrographs, would reveal the exciting sources. Similarly, the  $\sigma$ 
{\it vs} $\lambda_c$ new diagram picks up the small bursts of stellar formation by tracing their lower intensity and 
slower expansion of their immediate surroundings. 

The stalactites detected in the $\sigma$ {\it vs} $\lambda_c$ diagram in the case of\, III~Zw~102\, and Mrk~370 
correspond to the ionized knots located in the outskirts of the galaxies and their
spread in $\sigma$ result from the H {\small II} regions expansion 
into a low density medium \citep[see][]{Franco90}.

On the whole, the new diagnostic diagrams here presented provide   the 
possibility of inferring from 3D spectroscopy,  the magnitude of the 
star formation activity of distant galaxies, even if these are not
spatially resolved.

\acknowledgments 
IMD acknowledges the FPI grant (FP-2001-2506) of the Spanish Government through the 
co\-lla\-bo\-ra\-tion of the project AYA2004-08260-C03-01 (ESTALLIDOS,
\\http://www.iac.es/project/GEFE/es\-ta\-lli\-dos).
AVM acknowledges the the Russian Foundation for Basic Research (project
05-02-16454).
This work is partly based on observations carried out at the 6m telescope of the Special Astrophysical 
Observatory of the Russian Academy of Sciences, operated under the financial support of the Science Department 
of Russia (registration number 01-43),
and it has been partly supported by the ESTALLIDOS Project (AYA2004-08260-C03-01) and the grant
AYA 2004-02703 from the Spanish Ministerio de Educaci\'on y Ciencia.


\begin{deluxetable}{lccccccc}
\tablecaption{Log of the Observations \label{logs}}
\tabletypesize{\small}
\tablehead{
\colhead{Galaxy} & \colhead{RA(2000)} & \colhead{DEC(2000)} & \colhead{Linear Scale(pc/$''$)}& \colhead{t(s)} &  
\colhead{PSF($''$)} &\colhead{V$_{HI}$(km s$^{-1}$)} & \colhead{D(Mpc)} \\
\colhead{(1)} & \colhead{(2)} & \colhead{(3)}& \colhead{(4)}& \colhead{(5)}& \colhead{(6)} & \colhead{(7)} & \colhead{(8)}}

\startdata
Mrk370	 &  02\hour40\mint29\fs0 & +19\arcdeg17\arcmin50\arcsec   &   52 & 4680 & 1.7 &  790$^\dag$ & 10.85\\
IIIZw102 &  23\hour20\mint30\fs1 & +17\arcdeg13\arcmin32\arcsec   &  110 & 6480 & 1.8 & 1626$^\dag$ & 22.71\\
Mrk324	 &  23\hour26\mint32\fs8 & +18\arcdeg15\arcmin59\arcsec   &  108 & 4320 & 1.6 & 1600$^\dag$ & 22.43\\
\enddata
\tablenotetext{\dag}{\small Systemic velocity, from H {\scriptsize I} observations, corrected to the Local Group velocity centroid
taken from \cite{Thuan81}.}
\end{deluxetable}

\end{document}